\documentclass[prb,aps,twocolumn,superscriptaddress,showpacs]{revtex4}
\usepackage{graphicx}
\usepackage{hyperref}

\begin{document}

\title{Coupled-mode theory for periodic side-coupled microcavity and photonic crystal
structures}
\author{Philip Chak}
\affiliation{Department of Physics and Institute for Optical
Sciences, University of Toronto, Ontario, Canada M5S 1A7}

\author{Suresh Pereira}
\affiliation{ Groupe d'Etude des Semiconducteurs, Unit\'e Mixte de
Recherche du Centre National de la Recherche Scientifique 5650,
Universit\'e Montpellier II, 34095 , Montpellier, France }

\author{J.E. Sipe}
\affiliation{Department of Physics and Institute for Optical
Sciences, University of Toronto, Ontario, Canada M5S 1A7}

\date{\today}

\begin{abstract}
We use a phenomenological Hamiltonian approach to derive a set of
coupled mode equations that describe light propagation in waveguides
that are periodically side-coupled to microcavities. The structure
exhibits both Bragg gap and (polariton like) resonator gap in the
dispersion relation. The origin and physical significance of the two
types of gaps are discussed. The coupled-mode equations derived from
the effective field formalism are valid deep within the Bragg gaps
and resonator gaps.
\end{abstract}


\maketitle

\affiliation{Department of Physics, University of Toronto,
Toronto, Ontario, Canada M5S 1A7}

\affiliation{Centre d'optique, photonique et laser, Universit\'{e}
Laval, Sainte-Foy, Qu\'{e}bec (Canada) G1K 7P4}

\affiliation{Department of Physics, University of Toronto,
Toronto, Ontario, Canada M5S 1A7}

\section{Introduction}

In the past several years the linear and nonlinear properties of
side-coupled waveguiding structures have attracted the attention of
many researchers \cite{LittlePaper}$^{-}$\cite{WaksPaper}. These
structures consist of one or more waveguiding elements in which
forward and backward propagating waves are \textit{indirectly}
coupled to each other \textit{via} one or more mediating resonant
cavities. Perhaps the most common proposals for realizing these
structures involve photonic crystal (PC)\ waveguides with defect
modes slightly displaced from the waveguiding region (Fig. 1a, left)
\cite{HausPaper1}$^{,}$\cite{YarivPaper1}, or micro-ring resonator
structures in which two channel waveguides are side-coupled to
micro-ring resonators (Fig. 1a, right) \cite{InvitedPaper}. In the
PC structure the forward and backward propagating modes within the
waveguide are coupled \textit{via} the defect; for the micro-ring
structure, the forward going mode in the lower (upper) channel
waveguide is coupled, \textit{via} the micro-ring, to the backward
going mode in the upper (lower) channel. The linear and nonlinear
properties of both types of structures have been studied
\cite{HausPaper1}$^{,}$\cite{YarivPaper1}$^{-}$\cite{SoljacicPaper}.

The electromagnetic properties of these structures can be accurately
determined in great detail using numerically intensive methods such as
finite-difference time-domain (FDTD) simulations \cite{FDTDbook}. An
analysis in terms of Wannier functions can substantially reduce computation
time for the PC structure \cite{KurtPaper1}, but the numerical problem
remains daunting. In particular, full FDTD calculations of the micro-ring
structures have to date been confined to two-dimensional analogs of the
actual structures of interest \cite{FDTDbook}. Furthermore, direct numerical
simulation, while valuable for design purposes, offers little insight into
the physics of the structures. Consequently, semi-analytical techniques,
such as the scattering-matrix approach of S. Fan \textit{et al}.\cite
{HausPaper1} and Yong Xu \textit{et al.}\cite{YarivPaper1}, have been
proposed. Using these techniques the optical properties of side-coupled
structures can be understood in terms of the interactions between a small
number of modes.

In this paper we concentrate our attention on \textit{periodic}, side
coupled structures (Fig. 1b). Our primary objective is to derive coupled
mode equations (CME) that describe pulse propagation in such structures.
Coupled mode theory has long been used as an effective design tool for
grating structures where forward and backward propagating waves are \textit{%
directly} coupled \textit{via} an index grating \cite{YarivQEBook}.
In directly coupled structures, it is well known that a Bragg gap
opens in the dispersion relation of the structure when the phase
accumulated
in one round trip through a period of the grating is an integer multiple of $%
2\pi $, so that the slight reflections that are incurred due to the
grating are coherently enhanced. Structures possessing a Bragg gap
have found a variety of uses, such as dispersion compensation
\cite{EggletonPaper} and wavelength division multiplexing
\cite{GilesPaper}. In the side-coupled structure the Bragg feedback
mechanism, and hence the Bragg gap, does exist, although it is now
mediated by the coupling cavity. However, there is also a second
type of gap: a \textit{resonator} gap, which is associated with the
resonance frequencies - and therefore the geometry - of the
mediating cavity. For the micro-ring resonator structure the
interpretation of this gap is straightforward: when the phase
accumulated in a round-trip through the micro-ring resonator is an
integer multiple of $2\pi $, then the coupling between the forward
and backward going waves is resonantly enhanced. Of these two gaps,
the resonator gap is perhaps the more important, because it exhibits
a deep transmission dip seen even in a structure with only one unit
cell.

Because side-coupled structures exhibit both Bragg and resonator
gaps, it is to be expected that a CME description of optical pulse
propagation will be more complicated than in Bragg gratings. The CME
for Bragg gratings involve two fields (forward and backward going)
interacting \textit{via }a coupling coefficient. For side-coupled
structures, the most interesting situation is when a resonator gap
lies near one of the Bragg gaps, and we show in this paper that the
relevant CME then involves three fields: a cavity field and forward
and backward going fields.

We derive our CME\ using a phenomenological Hamiltonian approach, which
distills the essential physical interactions of the structure, and hence
provides a simple physical picture of optical interactions. We build the
fields in our CME\ as Fourier superpositions of the modes in the
Hamiltonian. Hence, our CME\ are derived for infinite, periodic structures
in which the coupling to each cavity is the same. Nevertheless, we show that
our CME can be generalized to describe finite, apodized structures, in which
the coupling (but not the period) varies from cavity to cavity. Therefore,
the CME can be used to describe finite structures with only a small number
of cavities. Indeed, the general Hamiltonian approach we advocate can be
applied even to structures with only one or two cavities, if the formalism
we introduce in Sec. II is extended to a discrete number of (not necessarily
identical) cavities. In both discrete and periodic scenarios, the
Hamiltonian approach exhibits the similarities of the optical dynamics of
these artificially structured materials to more traditional problems in
solid state physics. As well, it allows for an easy quantization of the
description to address the quantum optics of these structures. We plan to
turn to this, as well as the direct derivation of our phenomenological
Hamiltonian from the underlying electrodynamics, in future publications.

The present paper is organized as follows. In Sec. II\ we describe the
Hamiltonian model for a system with a single microresonator, investigate the
transmission/reflection spectrum of the structure, and indicate how the
parameters in our phenomenological Hamiltonian can be set from more common
models of cavity resonators. In Sec. III we discuss how the Hamiltonian can
be used to model a periodic waveguide-resonator structure. We then discuss
methods of reducing the number of fields and interactions in our Hamiltonian
while retaining the basic physics. In Sec. IV we derive the coupled mode
equations in terms of effective fields built as Fourier superpositions of
the modes in the Hamiltonian of Sec. III, and we show how to modify these
CME to describe finite, apodized structures. In Sec. V we conclude.

\section{Hamiltonian model and transmission for a single cavity structure}

\begin{figure}[t]
\includegraphics[width=8cm,height=10cm,angle=270] {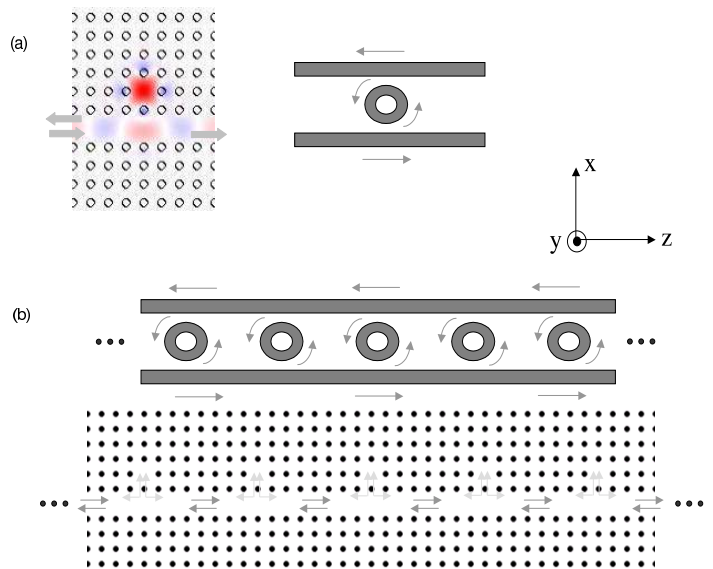}
\caption{ (1a) Waveguide-resonator structure containing (left)
photonic crystal microcavity and (right) micro-ring resonator. On
the right the micro-ring resonator is coupled to two waveguides,
with the forward (backward) propagating light in the lower (upper)
waveguide; On the left the structure contains dielectric rods is
embedded in air, the singly degenerate microcavity is coupled to the
photonic crystal waveguide, formed by removing a row of rods in the
photonic crystal. (1b) Periodic waveguide resonator structure
containing (top) microring resonator and (bottom) photonic crystal
microcavity. }
\end{figure}

In this section we construct a Hamiltonian model for a structure in which
forward and backward propagating waves are indirectly coupled to each other
\textit{via} a cavity centred at $z=z_{0}$. We will focus on classical
optics here, but because its easy generalization to quantum optics is one of
the strengths of this approach, we adopt a quantum notation and, for the
classical Poisson bracket $\left\{ ..,..\right\} $, we write $(i\hbar
)^{-1}\left[ ..,..\right] $; we also use $^{\dagger }$ to indicate complex
conjugation. We will also often speak of operators rather than variables,
especially when it makes the physics more clear. For example, we introduce $%
a_{k}^{\dagger }$ and $c_{k}^{\dagger }$ as creation operators for
photons propagating with wavenumber $k$ in the forward and backward
direction respectively. Because $k>0$ ($k<0$) indicates that the
photons are propagating in the forward (backward) direction,
$a_{k}^{\dagger }$ exists for $k>0$ and $c_{k}^{\dagger }$ for
$k<0$. For a given $k$, the energy in these fields is $\hbar \omega
_{k}a_{k}^{\dagger }a_{k}$ and $\hbar \omega _{k}c_{k}^{\dagger
}c_{k}$, with $\omega _{k}=c\left| k\right| /n$, where $c$ is the
speed of light in a vacuum, and $n$ is a constant effective index,
equal for the forward and backward propagating waves. By ignoring
the frequency dependence of $n$ we are neglecting the underlying
material dispersion within the waveguides; we discuss the validity
of this approximation after equation (\ref{2_Ham_Heis}) below. To
describe light in the cavity, we define a creation operator
$b^{\dagger }$, and identify the energy in the field as $\hbar
\omega _{0}b^{\dagger }b$, where $\omega _{0}$ is the resonant
frequency of the cavity. For the micro-ring resonator structure of
Fig. 1a (right), the $a_{k}^{\dagger }$ and $c_{k}^{\dagger }$ could
represent creation operators for light propagating in the forward
direction in the lower waveguide and the backward direction in the
upper waveguide, while $b^{\dagger }$ could represent the field
circulating in the counter-clockwise direction in the micro-ring
resonator. Our notation implies that the two waveguides have a
common mode index $n$, but this could
easily be generalized. For the PC structure of Fig. 1a (left), the $%
a_{k}^{\dagger }$ and $c_{k}^{\dagger }$ would represent creation operators
for light propagating in the forward and backward direction in a waveguide
mode of the PC waveguide, and $b^{\dagger }$ would represent the creation
operator for the field inside the single mode defect. Regardless of their
interpretation, the operators satisfy the commutation relations
\begin{eqnarray}
\left[ a_{k},a_{k^{\prime }}^{\dagger }\right] &=&\delta \left( k-k^{\prime
}\right) ,  \nonumber \\
\left[ c_{k},c_{k^{\prime }}^{\dagger }\right] &=&\delta \left( k-k^{\prime
}\right) ,  \nonumber \\
\left[ b,b^{\dagger }\right] &=&1,  \label{2_Ham_Comm}
\end{eqnarray}
with all other commutation relations vanishing. Assuming that no light
couples directly between the propagating modes governed by $a_{k}^{\dagger }$
and $c_{k}^{\dagger }$, but that light can couple from these modes to the
cavity, we use the following model Hamiltonian for the system \cite
{HausPaper1}$^{,}$\cite{YarivPaper1}:
\begin{equation}
H\mathbf{=}H_{o}+H_{coupling},  \label{2_Ham_eqn1} \\
\end{equation}
where
\begin{eqnarray}
H_{o} &=&\int_{0}^{\infty }dk\hbar \omega _{k}a_{k}^{\dagger
}a_{k}+\int_{-\infty }^{0}dk\hbar \omega _{k}c_{k}^{\dagger }c_{k}+\hbar
\omega _{o}b^{\dagger }b,  \nonumber  \label{2_Ham_eqn2} \\
&& \\
H_{coupling} &=&-\hbar \int_{0}^{\infty }\xi _{k}\left[ a_{k}^{\dagger
}be^{-ikz_{0}}+b^{\dagger }a_{k}e^{ikz_{0}}\right] dk  \nonumber \\
&&-\left( -1\right) ^{q}\hbar \int_{-\infty }^{0}\xi _{-k}\left[
c_{k}^{\dagger }be^{-ikz_{0}}+b^{\dagger }c_{k}e^{ikz_{0}}\right]
dk. \nonumber \\
\label{2_Ham_eqn3}
\end{eqnarray}
The quantities $\xi _{k}$ and $\left( -1\right) ^{q}\xi _{-k}$ characterize
the strength of the coupling between cavity field and waveguide fields,
propagating in the forward and backward direction; $q$ is an integer that
depends on the symmetry of the cavity mode \cite{YarivPaper1}. Note that
except for the factor $(-1)^{q}$ our notation implies that the coupling to
forward and backward propagating waveguide modes is identical. In the
micro-ring structure, for example, this means that we assume equal coupling
to the two waveguides; generalization of this is straightforward, but for
simplicity we will not do it here. The time evolution of the operators is
given by the Heisenberg equations of motion
\begin{equation}
i\hbar \frac{dO}{dt}=\left[ O,H\right] ,  \label{2_Ham_Heis}
\end{equation}
where $O$ is any operator.

In writing down (\ref{2_Ham_eqn1}), (\ref{2_Ham_eqn2}) and (\ref{2_Ham_eqn3}%
) we have implicitly assumed that the cavity supports only one mode,
with resonant frequency $\omega _{0}$, and that the waveguides guide
light in only a single spatial mode profile. Strictly speaking, of
course, neither of these assumptions is valid. In general, cavities
support more than one mode, oscillating at one or more resonance
frequencies, and for sufficiently high frequencies a waveguide will
support multiple transverse modes. However, we are primarily
interested in the physics of these structures for frequencies at or
near a specific resonant frequency $\omega _{0}$. We then assume
that within this frequency range only one resonance of the cavity
exists or, alternatively, that only a single mode of a multi-mode
cavity is excited, and that the waveguides of the structure are
single mode. Furthermore, we assume that the underlying material or
modal dispersion of the structure is negligible within the frequency
range of interest. For our purposes, the inclusion of material
dispersion would lead to quantitative, but not qualitative changes.

In Appendix $1$ we show that our Hamiltonian formulation leads to a
Lorentzian transmission and reflection across the cavity for frequencies in
the vicinity of $\omega _{0}$:
\begin{eqnarray}
t\left( \omega \right) &\simeq &\frac{-i\Delta }{\gamma -i\Delta },
\label{2_trans} \\
r\left( \omega \right) &\simeq &\left( -1\right) ^{q}\left( \frac{\gamma }{%
\gamma -i\Delta }\right) ,  \label{2_refl}
\end{eqnarray}
where $\gamma = 2\pi n \xi _{\tilde{\omega}_{0}}^{2} /c$,
and $\xi _{\tilde{%
\omega}_{o}}$ is the coupling coefficient between the cavity and
waveguides evaluated at $k= \tilde{\omega}_{0}\equiv n\omega
_{0}/c$, and where $\Delta =\left( \omega -\omega
_{0}-\alpha \left( \omega \right) \right) $ characterizes the
detuning from the renormalized resonance frequency $\omega
_{0}+\alpha \left( \omega \right) $ . An expression for the quantity
$\alpha \left( \omega \right) $ is given in Appendix $1$. For our
structures of interest $\alpha \left( \omega \right) $ is
sufficiently small that $\omega -\omega _{0}-\alpha \left( \omega
\right) \simeq \omega -\omega _{0}$ to a good approximation.

The transmission and reflection coefficients in (\ref{2_trans}),(\ref{2_refl}%
) are of precisely the form that follows from simple transfer matrix models
of resonant cavities or ring resonators \cite{YarivPaper1}$^{,}$\cite
{InvitedPaper}. In the latter structure, for example, the coupling of the
cavity to the waveguides is described by self-coupling and cross-coupling
coefficients $\sigma $ and $\kappa $ respectively, which in a simple case
(where the coupling is assumed to occur at the point of smallest separation)
are real and satisfy $\sigma ^{2}+\kappa ^{2}=1$. Comparing the transmission
and reflection coefficients found there with (\ref{2_trans}),(\ref{2_refl}),
we find that they become equivalent if we put
\begin{equation}
\gamma = \frac{c}{2 \pi \bar{n} R} \left( \frac{1-\sigma^{2}}{\sigma^{2}} \right)
\label{2_gamma_rel}
\end{equation}
where $\bar{n}$ and $R$ are the effective index and radius of the resonator
respectively. Thus if a given resonator is parameterized by $\sigma $ and $%
\kappa $, as well of course by the resonance frequency $\omega
_{0}$, then relation (\ref{2_gamma_rel}) allows one to determine the
effective coupling coefficient $\xi _{\tilde{\omega_{0}}}$ and thus
set what will be, as we will see, the crucial elements in the
phenomenological Hamiltonian (\ref{2_Ham_eqn1}). The appropriate
values of $\sigma \,$and $\kappa $ for a single resonator could be
determined by experiment, or directly calculated from the underlying
channel and resonator geometries, as discussed by Waks and Vuckovic
\cite{WaksPaper}.

A typical spectrum for a single cavity structure is shown in Fig. 2. On
resonance, the reflection induced by the cavity reaches 100\% (albeit only
for a single wavelength), and remains significant as long as the detuning, $%
\Delta $, is on the order of $\gamma$. The width of the spectrum is
dictated by $\gamma$, and the larger the coupling to the cavity, the
broader the resonance. In physical terms, this means that as the waveguides
are brought \textit{closer} to the cavity of Fig. $1a$, the resonance width
increases.

\begin{figure}[t]
\includegraphics[width=5cm,angle=270] {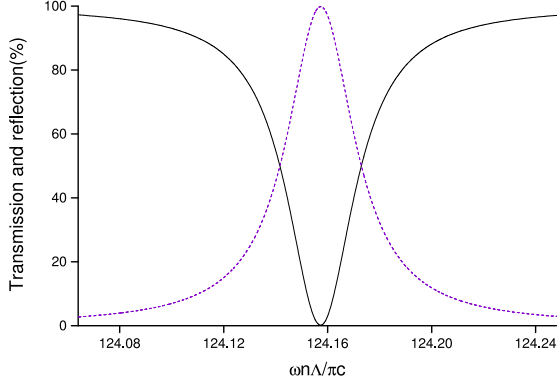}
\caption{ Transmission (solid line) and reflection (dotted line)
spectrum for the one cell structure obtained using equations\
(\ref{2_trans}) and (\ref {2_refl}). The structure can demonstrate
100\% reflection and 0\% transmission when the frequency is matched
to the resonance frequency of the microresonator. For comparison
with later plots, the frequency is normalized with a distance
$\Lambda$, which we use as the distance between resonators when we
consider a periodic array. }
\end{figure}

\section{Hamiltonian for a periodic structure}

\smallskip We now generalize the single-cavity Hamiltonian to describe a
periodic structure, in which the forward and backward propagating modes are
coupled to an infinite series of periodically spaced cavities (Fig. 1b). We
assume that the resonators are not directly coupled to each other, although
of course they do couple indirectly \textit{via }the waveguides.
Generalizing the Hamiltonian (\ref{2_Ham_eqn1}) to include the periodic
sequence of resonators, we write

\begin{eqnarray}
H &=&\int_{0}^{\infty }dk\hbar \omega _{k}a_{k}^{\dagger
}a_{k}+\int_{-\infty }^{0}dk\hbar \omega _{k}c_{k}^{\dagger
}c_{k}+\sum_{l}\hbar \omega _{o}b_{l}^{\dagger }b_{l}  \nonumber
\label{4_Ham_extz} \\
&&-\hbar \sum_{l}\int_{0}^{\infty }dk\xi _{k}\left[ b_{l}^{\dagger
}a_{k}e^{ikz_{l}}+a_{k}^{\dagger }b_{l}e^{-ikz_{l}}\right]  \nonumber \\
&&-\left( -1\right) ^{q}\hbar \sum_{l}\int_{-\infty }^{0}\xi _{-k}dk\left[
b_{l}^{\dagger }c_{k}e^{ikz_{l}}+c_{k}^{\dagger }b_{l}e^{-ikz_{l}}\right] ,
\nonumber \\
&&
\end{eqnarray}
where $a_{k}^{\dagger }$ $\left( c_{k}^{\dagger }\right) $ are again the
creation operators for light propagating the forward (backward) direction.
The main difference between (\ref{4_Ham_extz}) and (\ref{2_Ham_eqn1})\ is
that we have now included a countably infinite number of resonators, each
with the same resonance frequency, $\omega _{0}$, and associated with the
creation operator $b_{l}^{\dagger }$, where $l$ indexes the resonator. The
resonators are evenly spaced at $z_{l}=l\Lambda $, which gives a fundamental
reciprocal lattice vector $G_{0}=2\pi /\Lambda $. The Hamiltonian (\ref
{4_Ham_extz}) can be re-written as
\begin{widetext}
\begin{eqnarray}
H &=&\sum_{G}\int_{B.Z.}dk\hbar \omega _{k+G}a_{k+G}^{\dagger
}a_{k+G}+\sum_{G}\int_{B.Z.}dk\hbar \omega _{k-G}c_{k-G}^{\dagger
}c_{k-G}+\sum_{l}\hbar \omega _{o}b_{l}^{\dagger }b_{l}  \nonumber \\
&&-\hbar \sum_{l}\sum_{G}\int_{B.Z.}dk\xi _{k+G}\left[ b_{l}^{\dagger
}a_{k+G}e^{i\left( k+G\right) z_{l}}+a_{k+G}^{\dagger }b_{l}e^{-i\left(
k+G\right) z_{l}}\right]   \nonumber \\
&&-\left( -1\right) ^{q}\hbar \sum_{l}\sum_{G}\int_{B.Z.}dk\xi
_{-k+G}\left[ b_{l}^{\dagger }c_{k-G}e^{i\left( k-G\right)
z_{l}}+c_{k-G}^{\dagger }b_{l}e^{-i\left( k-G\right) z_{l}}\right] ,
\label{4_Ham_redz}
\end{eqnarray}
\end{widetext}
where $\sum_{G}$ represents the summation over an infinite number of \textit{%
positive} reciprocal lattice vectors (with $G=0,G_{0},2G_{0},...$\textbf{)},
and where in the integrations we restrict the wavenumber $k$ to the first
Brillouin zone ($-G_{0}/2<k\leq G_{0}/2$); We sum only over the positive
reciprocal lattice vectors so that $a_{k+G}^{\dagger }$ and $%
c_{k-G}^{\dagger }$ retain their association with forward and backward
propagation modes respectively. The operators satisfy commutation relations
\begin{eqnarray}
\left[ a_{k+G},a_{k^{\prime }+G^{\prime }}^{\dagger }\right] &=&\delta
\left( k-k^{\prime }\right) \delta _{G,G^{\prime }},  \nonumber \\
\left[ c_{k-G},c_{k^{\prime }-G^{\prime }}^{\dagger }\right] &=&\delta
\left( k-k^{\prime }\right) \delta _{G,G^{\prime }},  \nonumber \\
\left[ b_{l},b_{l^{\prime }}^{\dagger }\right] &=&\delta _{l,l^{\prime }},
\label{4_comm_rel}
\end{eqnarray}
with all other commutators vanishing; the first two of these follow
immediately from (\ref{2_Ham_Comm}).\ Because the system is periodic, we can
identify a countably infinite set of \textit{Bragg frequencies} in (\ref
{4_Ham_redz}). These are the frequencies $\omega _{k\pm G}$ evaluated at $%
k=0 $ or $G_{0}/2$. Hence, since $\omega _{k\pm G}=c\left| k\pm G\right| /n$
for ring resonator structures, the $M^{th}$ Bragg frequency occurs at $%
\omega _{b}^{(M)}=M\left( cG_{0}/2n\right) $ (with $M\geq 0$ an
integer).

To simplify (\ref{4_Ham_redz}), we introduce the \textit{collective operator}
\begin{equation}
b_{k}=\sqrt{\frac{\Lambda }{2\pi }}\sum_{l}b_{l}e^{-ikz_{l}},
\label{4_Ham_coll_op}
\end{equation}
where $k$ is now a continuous variable that ranges over the first Brillouin
zone. In Appendix 2 we introduce this operator by first considering only
excitations of the resonators periodic over a length $L=N\Lambda $, and then
taking $N\rightarrow \infty $. We find in that limit
\[
\sum_{l}\hbar \omega _{0}b_{l}^{\dagger }b_{l} \rightarrow
\int_{B.Z.}dk\hbar \omega _{o}b_{k}^{\dagger }b_{k},
\]
and that
\[
\left[ b_{k},b_{k^{\prime }}^{\dagger }\right] =\delta \left( k-k^{\prime
}\right)
\]
for $k$ and $k^{\prime }$ in the first Brillouin zone, with all other
commutators vanishing. In terms of this collective operator the Hamiltonian (%
\ref{4_Ham_redz})\ becomes
\begin{widetext}
\begin{eqnarray}
H &=&\sum_{G}\int_{B.Z.}dk\hbar \omega _{k+G}a_{k+G}^{\dagger
}a_{k+G}+\sum_{G}\int_{B.Z.}dk\hbar \omega _{k-G}c_{k-G}^{\dagger
}c_{k-G}+\int_{B.Z.}dk\hbar \omega _{o}b_{k}^{\dagger }b_{k}  \nonumber \\
&&-\hbar \sum_{G}\int_{B.Z.}dk\Xi _{+k+G}\left[ b_{k}^{\dagger
}a_{k+G}+a_{k+G}^{\dagger }b_{k}\right] -\left( -1\right) ^{q}\hbar
\sum_{G}\int_{B.Z.}dk\Xi _{-k+G}\left[ b_{k}^{\dagger
}c_{k-G}+c_{k-G}^{\dagger }b_{k}\right] ,  \label{4_Ham_final}
\end{eqnarray}
\end{widetext}
where $\Xi _{\pm k\pm G}$ $\equiv \sqrt{\frac{2\pi }{\Lambda }}\xi
_{\pm k\pm G}$. In Table I we give typical values for parameters
characterizing side-coupled structures, and we use them in our
sample calculations below.  There and for the rest of this paper we
assume that the coupling $\Xi _{\pm k\pm G^{\prime }}$ is
approximately constant at wavevectors corresponding to frequencies
within our region of interest, and take $\Xi _{\pm k\pm G^{\prime
}}\approx \Xi $. This approximation is reasonable if the $G$ of
interest satisfy $G\ll \Delta k$, where $\Delta k$ is the range over
which the $\xi _{k}$ varies significantly. We can expect $\Delta
k\approx 2\pi /(1\mu m)$ for the structures of interest (see
Appendix 1), and since $G$ is at most a few times $G_{0}=2\pi
/\Lambda $ ($=2\pi /(32\mu m)$ from Table I), this inequality is
indeed satisfied.

The dispersion relation of the system can be determined by
traditional transfer matrix methods, using (\ref{2_trans}),
(\ref{2_refl}) for the transmission and reflection coefficients of a
single resonator. However, to see the connection with the coupled
mode equations we will derive, we consider determining the
dispersion relation directly from the Hamiltonian
(\ref{4_Ham_final}), by applying the Heisenberg equation of motion
to generate equations for the time derivatives of $a_{k+G}$,
$c_{k-G}$ and $b_{k}$. Assuming harmonic time dependence
$e^{-i\omega t}$ for the operators, we determine an expression for
$\omega $ as a complicated function of the countably infinite set of
$\omega _{\pm \left| k\right| \pm G}$, and the discrete value
$\omega _{0}$. Alternately (and equivalently) we can exhibit the
Hamiltonian in a matrix form (\ref{4_Ham_final})
\begin{equation}
H=\hbar \int_{BZ}dk\ \mathbf{f}_{k}^{\dagger }\cdot \mathsf{V}%
_{k}\cdot \mathbf{f}_{k},  \label{4_Ham_full_matrix}
\end{equation}
where
\begin{equation}
\mathbf{f}_{k}^{\dagger }=\left(
a_{k+G_{0}}^{\dagger},a_{k+2G_{0}}^{\dagger},...,c_{k-G_{0}}^{\dagger},c_{k-2G_{0}}^{\dagger},...
,b_{k}^{\dagger}\right)
,
\end{equation}
and $\mathsf{V}_{k}$ contains all of the interactions between the
$a_{k+G}$, $c_{k-G}$ and $b_{k}$. Then, by diagonalizing the
(infinite-dimensional) matrix $\mathsf{V}_{k}$ we can in principle
determine the dispersion relation of the
structure. In Fig. 3 we consider a typical uncoupled (in the limit where $%
\Xi = 0$) and coupled dispersion relation for the structure. The
dotted line shows the uncoupled dispersion relation, and the solid
line shows the dispersion relation of the coupled system, as
determined by the transfer matrix approach.

\begin{widetext}
\begin{table}
\caption{Parameters used for dispersion relation calculation.}
\begin{tabular}{|l|l|l|l|l|}
\hline Physical parameters & $\sigma =0.98$ & $\Lambda =32.0\mu m$ &
$n=3.0$ & $2\pi R=26.3\mu m$ \\
\hline Numerical parameters  & $\frac{\Xi}{c} =0.0023\mu m^{-1}$ &
$\frac{\omega _{0}n\Lambda }{\pi c}=124.156$ & $\frac{\omega
_{b}n\Lambda }{\pi c}=124.0$ & $\frac{\Xi n\Lambda }{\pi c}=0.07 \begin{tabular}{lllll}  \\  \end{tabular}$ \\
\hline
\end{tabular}
\end{table}
\smallskip
\end{widetext}

If one of the Bragg frequencies is close to the resonant frequency
$\omega _{0}$, then we show below that a truncation of the matrix
\textsf{V}$_{k}$ to three terms is a good approximation. The
restricted Hamiltonian that results is

\begin{widetext}
\begin{equation}
H \simeq \hbar \int_{B.Z.}dk\left[
\begin{array}{lll}
a_{k+G^{\prime }}^{\dagger } & c_{k-G^{\prime }}^{\dagger } &
b_{k}^{\dagger }
\end{array}
\right] \left[
\begin{array}{lll}
\omega _{k+G^{\prime }} & 0 & \Xi \\
0 & \omega _{k-G^{\prime }} & \left( -1\right) ^{q}\Xi
\\
\Xi  & \left( -1\right) ^{q}\Xi & \omega _{0}
\end{array}
\right] \left[
\begin{array}{l}
a_{k+G^{\prime }} \\
c_{k-G^{\prime }} \\
b_{k}
\end{array}
\right] .  \label{4_Ham_final_matrix}
\end{equation}
\end{widetext}

where $G^{\prime }$ is the reciprocal lattice vector associated with
the forward (backward) band that has $\omega _{k+G^{\prime }}$
$(\omega _{k-G^{\prime }})$ closest to $\omega _{0}$. Here we have
assumed that the resonant frequency is very close to a Bragg
frequency with its associated gap at the Brillouin zone centre, and
so $\omega _{G^{\prime }}=\omega _{-G^{\prime }}\equiv \omega _{b}$,
where $\omega _{b}$ is the Bragg frequency closest to the resonance
frequency$^{8}$.
We refer to eqn. 16 as the ``three mode model.''
Its validity near a resonance frequency for any particular structure
can be formally investigated by including the omitted terms in a
multiple scales analysis, or by simply comparing the dispersion
relation following from eqn. 16 with a full solution of the
dispersion relation using a transfer matrix approach. This is done
in Fig. 4, using the parameters in Table I as was done in Fig. 3.
In Fig. 4 we also plot the imaginary part of $k$ within the gaps.
Note that the exact solution and that from the three mode model are
in good agreement for the frequency range shown in Fig. 4. Such
agreement fails at other Bragg frequencies that are further from the
resonant gap, of course, since the three mode model (eqn. 16) only
contains the physics of the Bragg gap closest to $\omega _{0}$. It
is to frequencies near $\omega _{0}$ that we henceforth restrict
ourselves.

\begin{figure}[h]
\includegraphics[width=6cm,height=8cm,angle=270] {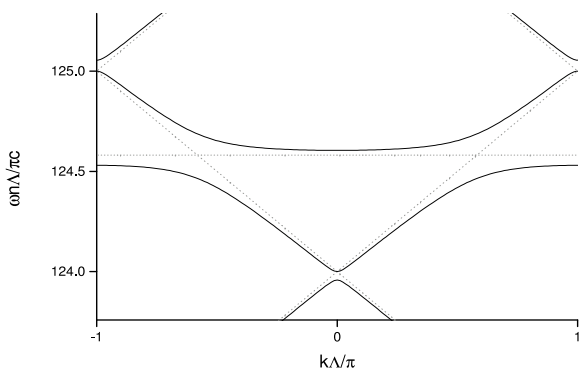}
\caption{ Typical dispersion relation for coupled microresonator
system as depicted in Fig. 1 (solid line). For comparison, the
dispersion relation of the system in the limit of no coupling
(dotted line) is also shown. The resonance frequency of the cavity
is given by $\omega _{0} n \Lambda/ \pi c = 124.58$ }
\end{figure}

\begin{figure}[t]
\includegraphics[width=16cm,height=8cm,angle=270] {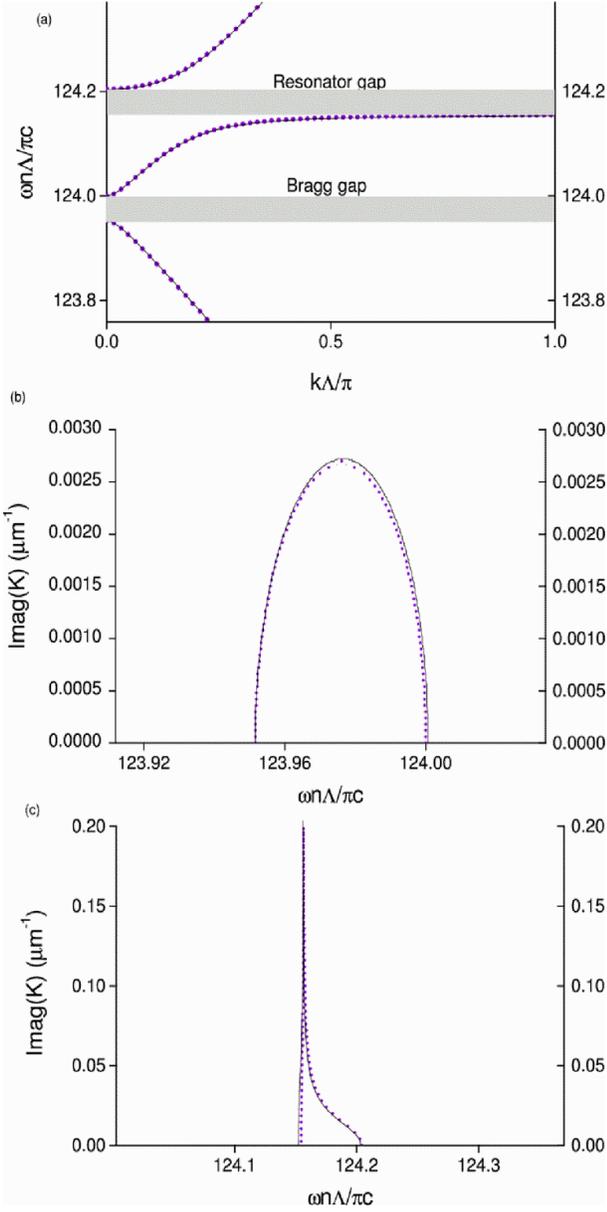}
\caption{ Dispersion relation obtained using the transfer matrix technique
(solid line) and the Hamiltonian in (\ref{4_Ham_final_matrix}) (circles).
(a) The real part of the dispersion relation. (b) The imaginary part of the
wavenumber for frequencies within the Bragg gap (c) The imaginary part of
the wavenumber for frequency within the resonator gap. }
\end{figure}

\section{Coupled-mode equations in the three-mode model}

\begin{figure}[b]
\includegraphics[width=13cm,height=8cm,angle=270] {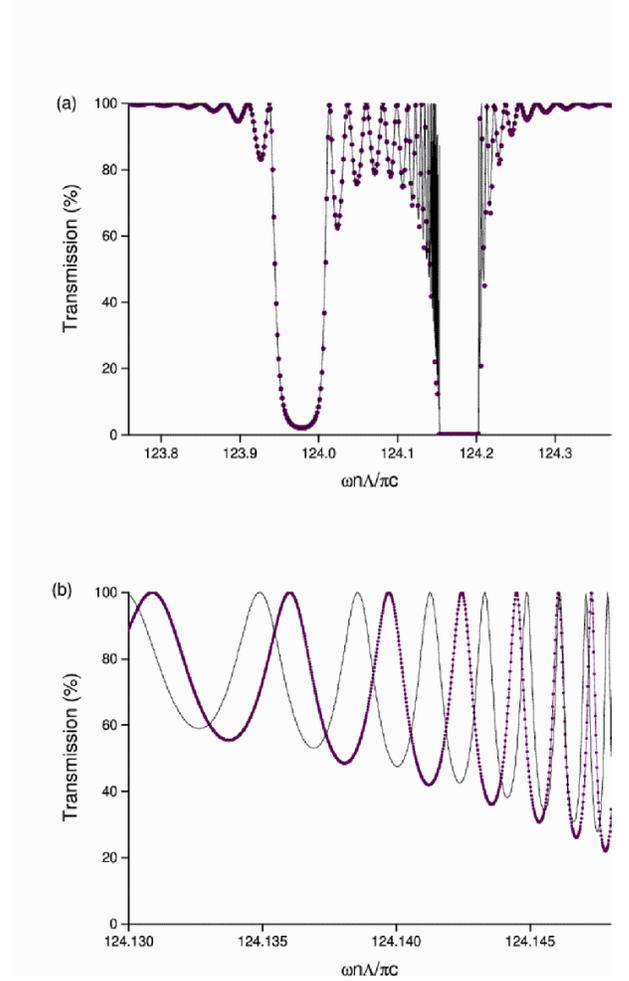}
\caption{ Transmission spectrum for finite structure that contains
30 cavities, using parameters depicted in Table I. (a) Solid line
represents the transmission spectrum obtained using coupled mode
equations and circles represents the transmission spectrum obtained
using transfer matrix. (b) Transmission spectrum in the vicinity of
the resonator gap using coupled mode equations (solid line) and
transfer matrix (solid line with circles).}
\end{figure}

In this section we derive a set of coupled-mode equations which describe
pulse propagation in the periodic structure, based on the three-mode
Hamiltonian (\ref{4_Ham_final_matrix}). We then demonstrate that although
these coupled mode equations are derived for an infinite periodic system
with equal coupling at each resonator, they can, with only slight
modifications, be used to describe finite systems with varying coupling at
each resonator. We start by defining effective fields in terms of the
amplitudes $a_{k+G^{\prime }}$, $c_{k-G^{\prime }}$ and $b_{k}$:
\begin{eqnarray}
g_{+}\left( z,t\right)  &=&\int_{B.Z.}\frac{dk}{\sqrt{2\pi }}a_{k+G^{\prime
}}e^{ikz},  \nonumber \\
g_{-}\left( z,t\right)  &=&\int_{B.Z.}\frac{dk}{\sqrt{2\pi }}c_{k-G^{\prime
}}e^{ikz},  \nonumber \\
b\left( z,t\right)  &=&\int_{B.Z.}\frac{dk}{\sqrt{2\pi }}b_{k}e^{ikz}.
\label{5_effective_fields}
\end{eqnarray}
where $G^{\prime }$ indexes the reciprocal lattice vector that is
retained within the three mode approximation. These fields can be
interpreted as a forward propagating field, a backward propagating
field, and the field distribution in the resonators respectively.
Using the definitions in (\ref {5_effective_fields}), the effective
fields satisfy the equal time commutation relations,
\begin{eqnarray}
\left[ g_{\pm }\left( z,t\right) ,g_{\pm }^{\dagger }\left( z^{\prime
},t\right) \right]  &=&\hat{\delta}\left( z-z^{\prime }\right)   \nonumber \\
\left[ b\left( z,t\right) ,b^{\dagger }\left( z^{\prime },t\right) \right]
&=&\hat{\delta}\left( z-z^{\prime }\right) ,  \label{5_eff_field_Comm}
\end{eqnarray}
with all other commutation relations vanishing. The function $\hat{\delta}%
(z-z^{\prime })$ is an \textit{effective delta function} such that $%
\int_{-\infty }^{\infty }f(z)\hat{\delta}(z-z^{\prime })dz=f(z^{\prime })$
when the function $f(z)$ has its wavenumber restricted to the first
Brillouin zone of the system. In terms of the effective fields, the
Hamiltonian in (\ref{4_Ham_final_matrix}) becomes
\begin{eqnarray}
H &=&\hbar \omega _{b}\int dzg_{+}g_{+}^{\dagger }+i\frac{\hbar c}{%
2n}\int dz\left( \frac{\partial g_{+}^{\dagger }}{\partial z}%
g_{+}-g_{+}^{\dagger }\frac{\partial g_{+}}{\partial z}\right)   \nonumber \\
&+&\hbar \omega _{b}\int dzg_{-}g_{-}^{\dagger }-i\frac{\hbar c}{2n}\int
dz\left( \frac{\partial g_{-}^{\dagger }}{\partial z}g_{-}-g_{-}^{\dagger }%
\frac{\partial g_{-}}{\partial z}\right)   \nonumber \\
&+&\hbar \omega _{0}\int dzbb^{\dagger }-\hbar \Xi \int dz\left( b^{\dagger
}g_{+}+c.c.\right)   \nonumber \\
&-&\left( -1\right) ^{q}\hbar \Xi \int dz\left( b^{\dagger
}g_{-}+c.c.\right)   \label{5_eff_field_Ham}
\end{eqnarray}
where $\omega _{b}$ denotes the Bragg frequency centered at the
Brillouin zone center and closest to $\omega _{0}$
\cite{Bragg_definition}. Using the Heisenberg equations of motion
for the effective fields, we obtain the coupled equations
\begin{eqnarray}
\left( \frac{\partial }{\partial t}+\frac{c}{n}\frac{\partial }{\partial z}%
\right) g_{+}\left( z,t\right)  &=&-i\omega _{b}g_{+}\left( z,t\right) +i\Xi
b\left( z,t\right) ,  \nonumber  \label{5_CME_1} \\
\left( \frac{\partial }{\partial t}-\frac{c}{n}\frac{\partial }{\partial z}%
\right) g_{-}\left( z,t\right)  &=&-i\omega _{b}g_{-}\left( z,t\right)
+i\left( -1\right) ^{q}\Xi b\left( z,t\right) ,  \nonumber \\
\frac{\partial }{\partial t}b\left( z,t\right) =-i\omega _{o}b\left(
z,t\right)  &+&i\Xi g_{+}\left( z,t\right) +i\left( -1\right) ^{q}\Xi
g_{-}\left( z,t\right) .  \nonumber \\
&&
\end{eqnarray}
One can obtain the dispersion relation directly from (\ref{5_CME_1})
by assuming that each field is a plane wave $e^{ikz-i\omega t}$,
with $k$ restricted to the first Brillouin zone. The results are
equivalent to those in Fig.4, obtained by diagonalizing
(\ref{4_Ham_final_matrix}).

Although the CME (\ref{5_CME_1}) were derived assuming an infinite medium,
they can be used to describe a structure where the coupling constant $\Xi $
varies slowly over a distance on the order of the spacing between the
resonators. A multiple scale analysis can be used to identify this limit and
corrections to it. A more striking inhomogeneous structure is one beginning
with a region where there are no resonators, followed by a length $L$ over
which resonators are placed with an equal spacing and equal coupling to the
channel(s), followed by a region where again there are no resonators. A
simple model for such a region would be to use the equations (\ref{5_CME_1}%
), but replacing $\Xi $ with a position dependent coupling constant $\left[
\theta \left( z\right) -\theta \left( z-L\right) \right] \Xi ,$ where $%
\theta $ is the usual step function. It can be easily seen that this model
formally violates our assumptions. Consider, for example, fields with a
stationary time dependence, so $g_{+}\left( z,t\right) =g_{+}\left( z\right)
e^{-i\bar{\omega}t}$, and similarily for all other fields. Then the first
equation gives
\begin{widetext}
\begin{equation}
-i\bar{\omega}g_{+}\left( z\right) +\frac{c}{n}\frac{\partial }{\partial z}%
g_{+}\left( z\right) =-i\omega _{b}g_{+}\left( z\right) +i\left[ \theta
\left( z\right) -\theta \left( z-L\right) \right] \Xi b\left( z,t\right) ,
\label{5_CME_ext_2}
\end{equation}
\end{widetext}
where in fact the factor $\left[ \theta \left( z\right) -\theta \left(
z-L\right) \right] $ could be omitted, since the third of (\ref{5_CME_1})
together with the position dependent coupling constant guarantees that $%
b\left( z\right) $ will only be nonzero in the region between $z=0$ and $z=L$%
. Note however that at $z=0$ and $z=L$ the equation
(\ref{5_CME_ext_2}) leads to a discontinuous $\partial g_{+}\left(
z\right) /\partial z$ if it is assumed that $g_{+}(z)$ is everywhere
continuous. This violates, of course, the
assumption that fields such as $g_{+}\left( z,t\right) $ are of the form  (%
\ref{5_effective_fields}).

Despite such a formal violation of our assumptions, this simple
model in fact gives a good description of the optical response of
a finite structure. To see this, consider first the fields
$g_{\pm}(z,t)$ within the structure.
It is clear from (\ref{5_CME_1}) that for a supposed frequency $%
\overline{\omega }$ there are two Bloch wavenumbers, which
equivalently follow
from (\ref{4_Ham_final_matrix}); they are given by $k\left( \bar{\omega}%
\right) =\pm \bar{k}$, where
\begin{equation}
\bar{k}=\frac{n}{c}\sqrt{\frac{\left( \Delta _{o}\Delta
_{1}-\Xi ^{2}\right) ^{2}-\Xi ^{4}}{\Delta _{0}^{2} }}.
\end{equation}
In the equation above $\Delta _{0}=\left(
\bar{\omega}-\omega _{0}\right) $ is the
detuning from the resonance frequency and $\Delta _{1}=\left( \bar{%
\omega}-\omega _{b}\right) $ is the detuning from the Bragg frequency that
lies closest to $\omega _{0}$. As a result, one can write the forward and
backward propagating effective fields, $g_{\pm }\left( z,t\right) $, as
\begin{eqnarray}
g_{\pm }\left( z,t\right)  &=&g_{\pm }\left( z\right)
e^{-i\bar{\omega}t}
\nonumber \\
g_{\pm }\left( z\right)  &=&g_{\pm }^{\left( 1\right)
}e^{i\bar{k}z}+g_{\pm }^{\left( 2\right) }e^{-i\bar{k}z},
\end{eqnarray}
Once $g^{(1)}_{+}$ and $g^{(2)}_{+}$ are set, $g^{(1)}_{-}$ and
$g^{(1)}_{-}$ are determined by the dispersion relation, or
equivalently (\ref{5_CME_1}). Hence there are only two independent
constants. Outside the structure ($\Xi = 0$) there are also two
independent constants in each of the regions $z < 0$ and $z > L$,
but the solution of (\ref{5_CME_1}) is simpler. There it takes the
form
\begin{eqnarray}
g_{\pm}(z,t) &=& g_{\pm}(z)e^{-i\bar{\omega}t}
\nonumber \\
g_{+}(z) &=& g_{+}e^{iqz} \nonumber \\
g_{-}(z) &=& g_{-}e^{-iqz}, \nonumber
\end{eqnarray}
where $ g_{+} $, $ g_{-} $ are independent and $q=\bar{\omega}n/c$.
For $z < 0$ we denote the constants by $g^{<}_{+}$ and $g^{<}_{-}$,
and for $z > L$ we denote them by $g^{>}_{+}$ and $g^{>}_{-}$. Now
we consider the boundary condition at $z = L$, and note that
since no field is incident from $z> L$, we have $g^{>}_{-}=0$;
an incident field is specified by
$g^{<}_{+}$. Our independent unknowns are then $g^{<}_{-},
g^{>}_{+}$, and the constants $g^{(1)}_{+}$ and $g^{(2)}_{+}$ that
specify the field in the structure. We solve for these four unknowns
by requiring the continuity of $g_{\pm}(z)$ at $z=0$ and $z=L$. The
resulting transmittance of the structure can be written as

\begin{equation}
T\left( \omega \right) =\left| \frac{ g^{>}_{+}e^{iqL} }{g_{+}^{(1)}+g_{+}^{(2)}}%
\right| ^{2}  \label{5_finite_transmission}
\end{equation}
with
\begin{eqnarray*}
g_{+}^{\left( 1\right) } &=&\frac{e^{-i\bar{k}L}}{2}\left[ 1+\frac{\Xi ^{2}}{%
\bar{k}\Delta _{o} }\left( \frac{\Delta _{o}\Delta
_{1}}{\Xi ^{2}}-1\right) \right]g^{>}_{+}e^{iqL},  \\
g_{+}^{\left( 2\right) } &=&\frac{e^{i\bar{k}L}}{2}\left[ 1-\frac{\Xi ^{2}}{%
\bar{k}\Delta _{o} }\left( \frac{\Delta _{o} \Delta _{1} }{\Xi ^{2}}-1\right)
\right]g^{>}_{+}e^{iqL} .
\end{eqnarray*}

In Fig. 5 we compare the transmission spectrum of a two channel
micro-ring resonator structure with 30 cavities, calculated both
using the transfer matrix technique,$^{7}$ and using the coupled
mode equation result eqn. 24. Again we adopt the parameters of Table
I. Generally there is good qualitative agreement, with the main
features of the spectrum well described by the coupled mode equation
result (24), although as noted above it is being applied beyond its
strict range of applicability. An extension of this approach leads
to the use of the CME (20) to treat a finite structure where the
coupling constant $\Xi $ varies from one resonator to the next. To
describe this we simply allow $\Xi $ in (20) to adopt a
$z$-dependence,
\begin{widetext}
\begin{eqnarray}
\left( \frac{\partial }{\partial t}+\frac{c}{n}\frac{\partial }{\partial z}%
\right) g_{+}\left( z,t\right) &=& -i\omega _{b}g_{+}\left(
z,t\right)
+i\Xi \left( z\right) b\left( z,t\right) ,  \nonumber \\
\left( \frac{\partial }{\partial t}-\frac{c}{n}\frac{\partial }{\partial z}%
\right) g_{-}\left( z,t\right) &=& -i\omega _{b}g_{-}\left(
z,t\right) +i\left( -1\right) ^{q}\Xi \left( z\right) b\left(
z,t\right),  \nonumber
\\
\frac{\partial }{\partial t}b\left( z,t\right) = -i\omega
_{0}b\left( z,t\right) &+& i\Xi \left( z\right) g_{+}\left(
z,t\right) +i\left( -1\right) ^{q}\Xi \left( z\right) g_{-}\left(
z,t\right). \label{5_apod_CMT}
\end{eqnarray}
\end{widetext}

\begin{figure}[h]
\includegraphics[width=13cm,height=8cm,angle=270] {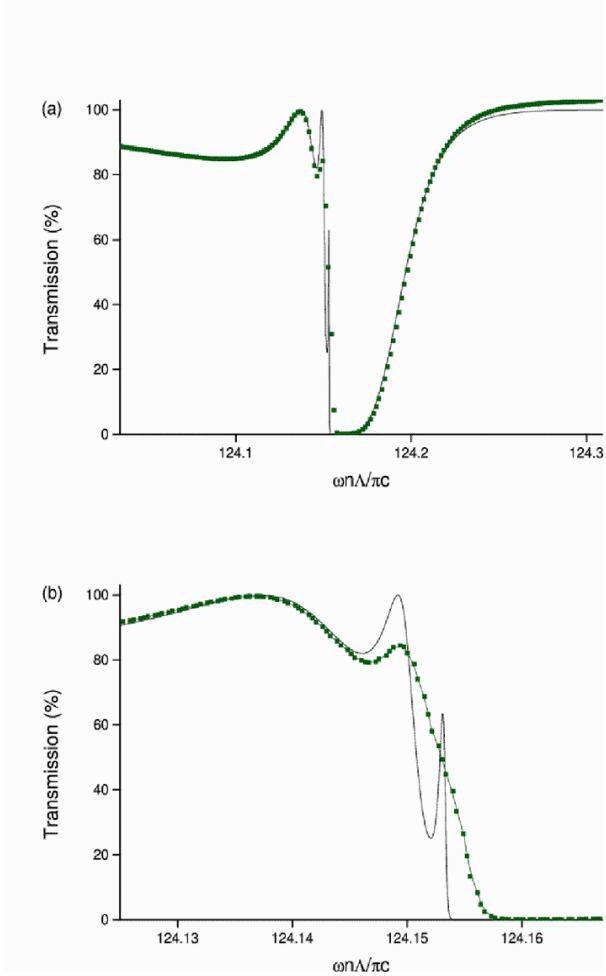}
\caption{Transmission spectrum for short, finite, apodized structure
with 5 unit cells. (a) Solid line represents the transmission
spectrum obtained using transfer matrix and squares represent
transmission spectrum obtained using coupled mode equations. (b)
Transmission spectrum in the vicinity of the resonator gap using
transfer matrix (solid line) and coupled mode equations (solid line
with squares).}
\end{figure}

In Fig. 6 we plot the transmission spectrum for a 5 cavity structure
apodized such that the cavities (from left to right) are
characterized by coupling constants $(\sigma _{1},...\sigma
_{5})=(0.993,0.986,0.98,0.986,0.993)$, corresponding to $(\Xi
_{1}\Lambda n/\pi c,..\Xi _{5}\Lambda n/\pi
c)=(0.0208,0.0287,0.0351,0.0287,0.0208)$. The transfer matrix
results is presented, as well as a very simple
application of the CME (25) using a piecewise uniform function to represent $%
\Xi $, where in the $n^{th}$ unit well we set $\Xi =\Xi _{n}$. Again
there is good qualitative agreement, although the CME are being
applied beyond their strict range of applicability. Besides the
difference between the CME and transfer matrix results with respect
to the Fabry-Perot type oscillations, as seen in Fig. 5, here the
CME solution also consistently overestimates the transmission on the
high-frequency side of the stop gap. This can be traced back to the
effects on the band curvature induced by the next highest Bragg gap,
which are implicitly included in the transfer matrix solution but
not in the CME calculation.

Finally, we note that while at least three coupled mode equations
are necessary to describe the kind of structures we consider here if
we deal with both their space and time dependence, if we instead
restrict ourselves to a stationary time dependence, $g_{\pm}\left(
z,t\right) =g_{\pm}\left( z\right) e^{-i\bar{\omega}t}$ and $b(z,t)
= b(z) e^{-i\bar{\omega}t}$, then in fact we can eliminate the
variable $b\left( z,t\right) $ and construct coupled mode equations
involving only $g_{+}\left( z,t\right) $ and $g_{-}\left( z,t\right)
$. They are

\begin{eqnarray}
\frac{\partial }{\partial z}g_{+}\left( z\right)  &=&i\nu \left( \omega
\right) g_{+}\left( z\right) +i\left( -1\right) ^{q}\mu \left( \omega
\right) g_{-}\left( z\right) ,  \nonumber  \label{5_Bragg_eqn} \\
\frac{\partial }{\partial z}g_{-}\left( z\right)  &=&-i\nu \left( \omega
\right) g_{-}\left( z\right) -i\left( -1\right) ^{q}\mu \left( \omega
\right) g_{+}\left( z\right) ,  \nonumber \\
&&
\end{eqnarray}
where
\begin{eqnarray}
\nu \left( \omega \right)  &=&\frac{n}{c}\left[ \frac{\Xi ^{2}}{\left(
\omega _{0}-\omega \right) }-\left( \omega _{b}-\omega \right) \right] ,
\nonumber \\
\mu \left( \omega \right)  &=&\frac{n}{c}\frac{\Xi ^{2}}{\left( \omega
_{0}-\omega \right) },  \label{5_Bragg_param}
\end{eqnarray}

These equations are valid for $\omega \neq \omega _{0}$. It is
well-known that a photonic band gap opens in the dispersion relation
described by these equations when $\left| \mu \left( \omega \right)
\right| \geq \left| \nu \left( \omega \right) \right| $,
\cite{ProgInOptics} and that the width of the gap is larger for
larger values of $\left| \mu \left( \omega \right) \right| $.
Consequently we see from these equations an analytic confirmation of
features that our dispersion relation display. Within our three mode
model, one edge of the resonator gap occurs at $%
\omega \rightarrow \omega _{0}$ (in which case $\nu $ and $\mu $
both diverge equally quickly and are hence equal in the limit as
$\omega $ approaches $\omega _{0}$), and one edge of the Bragg gap
occurs at $\omega \rightarrow \omega _{b}$, because then the second
term in the expression for $\nu \left( \omega \right) $ vanishes,
and $\nu \left( \omega _{b}\right) =\mu \left( \omega _{b}\right) $.

\section{Conclusion}

We have presented a phenomenological Hamiltonian description of
light propagation in side-coupled resonators. This formulation is
appealing in its
simplicity, since it captures the basic physics of the structures \textit{%
via }a set of readily understandably parameters. The most
interesting special case is perhaps where a resonator gap is close
to a Bragg gap, and at frequencies close to these gaps a three mode
model gives a good description of the dynamics of a periodic
structure of resonators. Coupled mode equations based on these
captures the dispersion relation even deep within the gaps, and a
naive extension of these equations to describe finite structures,
although not within the strict range of applicability of the model,
gives a good qualitative description.

A hallmark of the kind of approach we have taken here is the
connection of theoretically calculated or experimentally observed
parameters, such as the coupling coefficient $\sigma $, to the
parameters that appear in our phenomenological Hamiltonian. Such a
strategy is particularly amenable to the description of quantum and
nonlinear optical effects. The Hamiltonian description leads to
straightforward quantization, of course, and appropriate nonlinear
terms can easily be added to the Hamiltonian. In a previous study by
Grimshaw \textit{et al.}$^{22}$, it was shown that three nonlinear
coupled mode equations support stationary solitary wave solution in
the presence of Kerr nonlinearity. Numerical studies have indicated
that soliton-like waves exist in resonator structures. In future
work we plan to apply the approach we have detailed here to study
such field excitations, where a Hamiltonian framework provides the
ability to characterize conserved quantities in terms of the
symmetries of the nonlinear field theory.

\section{Acknowledgments}

This project was partly funded by the Natural Science and Engineering
Research Council (NSERC) of Canada. Philip Chak acknowledges financial
support from Photonic Research Ontario and an Ontario Graduate Scholarship.

\section{Appendix 1}

In this appendix we use the Hamiltonian (\ref{2_Ham_eqn1})\ to determine the
transmission properties of a single-cavity structure. These transmission
properties have been intensively studied using various methods such as
finite difference, time domain simulations\cite{FDTDbook}, and scattering
matrix techniques\cite{HausPaper1}\cite{YarivPaper1}, and it is well-known
that a Lorentzian function gives an excellent approximation to the response
of the structure. Here we show that our Hamiltonian also leads to a
Lorentzian spectrum. To discuss transmission and reflection, we assume that
there is a time-dependent source, $u\left( t\right) $, coupled to the
forward propagating modes at $z_{s}<z_{0}$. We therefore modify the
Hamiltonian (\ref{2_Ham_eqn1}) to include a source term:
\begin{equation}
H\mathbf{=}H_{o}+H_{coupling}+H_{source},  \label{3_Ham_eqn1}
\end{equation}
with
\begin{equation}
H_{source}=-\hbar \int_{0}^{\infty }\left[ a_{k}^{\dagger }u\left( t\right)
e^{-ikz_{s}}+a_{k}u^{*}\left( t\right) e^{ikz_{s}}\right] dk,
\label{3_Ham_eqn4}
\end{equation}
where $e^{ikz_{s}}$ accounts for the fact that the light is generated at $%
z=z_{s}$. Using the Hamiltonian (\ref{3_Ham_eqn1}) and the commutation
relations (\ref{2_Ham_Comm})\ in the\ Heisenberg equations of motion (\ref
{2_Ham_Heis}) we find
\begin{eqnarray}
a_{k}(t) &=&i\xi _{k}\int_{-\infty }^{t}b\left( t^{\prime }\right)
e^{-i\omega _{k}\left( t-t^{\prime }\right) }e^{-ikz_{o}}dt^{\prime }
\nonumber \\
&+&i\int_{-\infty }^{t}u\left( t^{\prime }\right) e^{-i\omega _{k}\left(
t-t^{\prime }\right) }e^{-ikz_{s}}dt^{\prime },  \nonumber \\
c_{k}(t) &=&i\left( -1\right) ^{q}\xi _{-k}\int_{-\infty }^{t}b\left(
t^{\prime }\right) e^{-i\omega _{k}\left( t-t^{\prime }\right)
}e^{-ikz_{o}}dt^{\prime },  \nonumber \\
\frac{db(t)}{dt} &=&-i\omega _{o}b(t)+i\int_{0}^{\infty }\xi
_{k}a_{k}(t)e^{ikz_{o}}dk  \nonumber \\
&+&i\left( -1\right) ^{q}\int_{-\infty }^{0}\xi _{-k}c_{k}(t)e^{ikz_{o}}dk.
\label{3_ak_eom}
\end{eqnarray}
where we have formally integrated the Heisenberg equations for $da_{k}/dt$
and $dc_{k}/dt$, so that both $a_{k}\left( t\right) $ and $c_{k}\left(
t\right) $ are expressed entirely in terms of $b\left( t\right) $ and $%
u\left( t\right) $. Using the expressions for $a_{k}\left( t\right) $ and $%
c_{k}\left( t\right) $ in the equation for $db/dt$, and expanding $b\left(
t\right) $ and $u\left( t\right) $ in terms of Fourier components,
\begin{eqnarray}
b\left( t\right) &=&\frac{1}{2\pi }\int_{-\infty }^{\infty }b\left( \omega
\right) e^{-i\omega t}d\omega ,  \nonumber \\
u\left( t\right) &=&\frac{1}{2\pi }\int_{-\infty }^{\infty }u\left( \omega
\right) e^{-i\omega t}d\omega ,  \label{3_Ham_bu_Four}
\end{eqnarray}
we obtain
\begin{equation}
b\left( \omega \right) =\left[ \frac{- 2\pi n \xi _{\widetilde{\omega }}/c }
{2 \pi n \xi _{%
\widetilde{\omega }}^{2}/c-i\Delta }\right] u\left( \omega \right) e^{i%
\widetilde{\omega }\left( z_{o}-z_{s}\right) },  \label{3_Ham_b}
\end{equation}
where $\Delta =\left( \omega -\omega _{o}+\alpha \left( \omega
\right) \right) $ and $\tilde{\omega} = \omega n/c$, with
\begin{eqnarray}
\alpha \left( \omega \right) =2\int_{0}^{\infty}\wp \left( \frac{ {\xi}%
_{k} ^{2}} {\frac{c}{n}k-\omega }\right) dk
\end{eqnarray}

describing the small shift in the resonance frequency of the cavity
due to the presence of the waveguide. To estimate the effect of
$\alpha (\omega)$, we assume $\xi_{k}$ takes a gaussian form in k
space with a peak centered at $k=\tilde{\omega_{0}}$. We take the
width of the gaussian profile to be about $~1\mu m^{-1}$, associated
with a typical length over which the coupling between the waveguide
and resonator is significant. Using this approximate form for
$\xi_{k}$ in the expression for $\alpha (\omega)$ and numerically
evaluating the integral, we have verified that $\alpha ( \omega ) $
is much smaller than the resonance frequency $\omega _{0}$ for
structures of interest. Note that in (\ref{3_Ham_b}) we
have switched our notation for wavenumber from $k$ to $\widetilde{\omega }%
=n\omega /c=\left| k\right| $ to stress that we are now considering
the frequency response of the structure. To determine the
transmission and reflection spectrum of the structure we define a
set of effective fields
\begin{eqnarray}
f_{+}\left( z,t\right) &=& \frac{1}{\sqrt{2 \pi}} \int_{0}^{\infty
}dka_{k}\left( t\right) e^{ikz},
\nonumber \\
f_{-}\left( z,t\right) &=& \frac{1}{\sqrt{2 \pi}} \int_{-\infty
}^{0}dkc_{k}\left( t\right) e^{ikz}. \label{3_Ham_eff_fields}
\end{eqnarray}
We then substitute the values (\ref{3_ak_eom}) for $a_{k}\left( t\right) $
and $c_{k}\left( t\right) $ in the effective fields (\ref{3_Ham_eff_fields}%
), and use the Fourier transforms (\ref{3_Ham_bu_Four}) of $b\left( t\right)
$ and $u\left( t\right) $ to simplify the integrals. We are specifically
interested in the following two quantities
\begin{widetext}
\begin{eqnarray}
\lim_{z\rightarrow \infty }f_{+}\left( z,t\right) &=&
\frac{i}{c} \int_{0}^{\infty }\left[ \frac{-i\Delta }{2 \pi n \xi _{\tilde{\omega%
} }^{2}/c-i\Delta }\right] u\left( \omega \right) e^{ik\left(
z-z_{s}\right)
}e^{-i\omega t}d\omega,  \label{3_field_eqn} \\
\lim_{z\rightarrow -\infty }f_{-}\left( z,t\right) &=&
\frac{i}{c}\left( -1\right) ^{q}\int_{0}^{\infty }\left[ \frac{2 \pi n \xi _{%
\tilde{\omega}}^{2}/c}{ 2 \pi n \xi _{\tilde{\omega}}^{2}/c-i\Delta
}\right] u\left( \omega \right) e^{ik\left( z+z_{s}\right)
}e^{i2\tilde{\omega} z_{o}}e^{-i\omega t}d\omega .  \nonumber
\end{eqnarray}
\end{widetext}
Note that in the absence of coupling we would have
\begin{widetext}
\begin{eqnarray}
\lim_{z\rightarrow \infty }f_{+}\left( z,t\right) &=& \frac{i}{c}
\int_{0}^{\infty }u\left( \omega \right) e^{ik\left( z-z_{s}\right)
}e^{-i\omega t}d\omega,  \label{3_field_eqn_no_coupling} \\
\lim_{z\rightarrow -\infty }f_{-}\left( z,t\right) &=& 0 \nonumber
\end{eqnarray}
\end{widetext}
The first (second) of the expressions in (\ref{3_field_eqn}) is the
transmitted
(reflected) field built as a superposition of the Fourier components of the source term, $%
u\left( \omega \right) $. We can therefore define the transmission and
reflection coefficients as
\begin{eqnarray*}
t\left( \omega \right) &=&\frac{-i\Delta }{2 \pi n \xi _{\tilde{\omega}}^{2}/c-i\Delta
}, \\
r\left( \omega \right) &=&\left( -1\right) ^{q}\frac{2 \pi n \xi _{\tilde{\omega}%
}^{2}/c}{2 \pi n \xi _{\tilde{\omega}}^{2}/c-i\Delta }.
\end{eqnarray*}
From these coefficients, it is clear that the cavity affects the
transmission/reflection of the structure when the detuning, $\Delta $, is on
the order of $2 \pi n \xi _{\tilde{\omega}}^{2}/c$. In the limit of very weak
coupling -- that is, when the value of $2 \pi n \xi _{\tilde{\omega}}^{2}/c$ is
approximately constant over a frequency range centered at $\omega_{0}$ and
spanning several multiples of $2 \pi n \xi _{\tilde{\omega}}^{2} /c$, then
the transmission and reflection are well approximated by a Lorentzian
lineshape
\begin{eqnarray}
t\left( \omega \right) &\simeq &\frac{-i\Delta }{\gamma -i\Delta }, \\
r\left( \omega \right) &\simeq &\left( -1\right) ^{q}\left( \frac{\gamma }{%
\gamma -i\Delta }\right) ,
\end{eqnarray}
where $\gamma \equiv 2 \pi n \xi _{\tilde{\omega}_{o}} /c$. This
condition yields $\gamma \ll c \Delta k/2n$; for our assumed $\Delta
k \simeq 2 \pi/(1 \mu m)$ this gives the requirement $\gamma \ll
300ps^{-1}$, which is met by typical values of $\gamma$ (see
equation (\ref{2_gamma_rel}) and Table I).

\section{Appendix 2}

In this appendix we build the continuous collective operator $b_{k}$ (\ref
{4_Ham_coll_op}) that applies for an infinite system of discrete resonators
by first considering only excitations that are periodic over a length $%
L=N\Lambda $, and then passing to the limit $N\rightarrow \infty $. In the
periodic case there are still an infinite number of resonators, but only $N$
of the $b_{l}$ are independent. Assuming $N$ is even, we can take them to be
\begin{equation}
l=-\frac{N}{2}+1,-\frac{N}{2}+2,...,\frac{N}{2}-1,\frac{N}{2}.
\label{lrange}
\end{equation}
We denote this range by $R$. For an $l$ outside $R_{l}$, we have $%
b_{l}=b_{l-pN}$ where $p$ is an integer such that $l-pN$ is within the range
(\ref{lrange}). If we now introduce discrete wavevectors $k_{m}=2\pi m/L$,
where
\begin{equation}
m=-\frac{N}{2}+1,-\frac{N}{2}+2,...,\frac{N}{2}-1,\frac{N}{2},
\label{mrange}
\end{equation}
(that is, $m\in R$) we can introduce Fourier amplitudes $\bar{b}_{m}$
according to
\begin{equation}
\bar{b}_{m}\equiv \frac{1}{\sqrt{N}}\sum_{l\in R}b_{l}e^{-ik_{m}z_{l}},
\label{FT}
\end{equation}
where $z_{l}=l\Lambda $. We then find immediately that
\[
b_{l}=\frac{1}{\sqrt{N}}\sum_{m\in R}\bar{b}_{m}e^{ik_{m}z_{l}},
\]
and that
\begin{equation}
\sum_{l\in R}b_{l}^{\dagger }b_{l}=\sum_{m\in R}\bar{b}_{m}^{\dagger }\bar{b}%
_{m},  \label{sumoverres}
\end{equation}
while
\[
\left[ \bar{b}_{m},\bar{b}_{m^{\prime }}^{\dagger }\right] =\delta
_{mm^{\prime }},
\]
for example, so
\[
\sum_{m^{\prime }}\left[ \bar{b}_{m},\bar{b}_{m^{\prime }}^{\dagger }\right]
=1
\]
or
\begin{equation}
\frac{2\pi }{L}\sum_{m^{\prime }}\left[ \sqrt{\frac{L}{2\pi }}\bar{b}_{m},%
\sqrt{\frac{L}{2\pi }}\bar{b}_{m^{\prime }}^{\dagger }\right] =1,
\label{dcomm}
\end{equation}
a form that we will presently find useful.

We now consider letting $N\rightarrow \infty $, with $L\rightarrow \infty $
such that $\Lambda $ is fixed. Then the range $R$ approaches all the
integers from $-\infty \,$to $+\infty $, while $k_{m}$ become more closely
spaced and approach a dense distribution of points ranging from $-\pi
/\Lambda $ to $\pi /\Lambda $; this is the first Brillouin zone, and we
denote it by $B.Z.$ In the usual way, then, we take
\begin{equation}
\frac{2\pi }{L}\sum_{m^{\prime }}\rightarrow \int_{B.Z.}dk^{\prime },
\label{sumtoint}
\end{equation}
and, if we introduce $b_{k}$ such that
\begin{equation}
\sqrt{\frac{L}{2\pi }}\bar{b}_{m}\rightarrow b_{k},  \label{distocon}
\end{equation}
where the $k$ in $b_{k}$ is first identified with $k_{m}$ but then allowed
to vary continuously as $N\rightarrow \infty $, from (\ref{dcomm}) we have
\[
\int_{B.Z.}dk^{\prime }\left[ b_{k},b_{k^{\prime }}^{\dagger }\right] =1,
\]
and so we can identify
\[
\left[ b_{k},b_{k^{\prime }}^{\dagger }\right] =\delta (k-k^{\prime }),
\]
for $k$ and $k^{\prime }$ within $B.Z.$ In this limit, using (\ref{sumtoint},%
\ref{distocon}), we find
\[
\sum_{l}b_{l}^{\dagger }b_{l}\rightarrow \int_{B.Z.}dk\,b_{k}^{\dagger
}b_{k}
\]
from (\ref{sumoverres}), where the integer $l$ now ranges from $-\infty $ to
$\infty $, and we recover (\ref{4_Ham_coll_op}) from (\ref{FT}).


\begin{references}
\bibitem{Bragg_definition} The Bragg frequency, $\omega_{b}$ is defined as
frequency satisfying $\omega_{b} = M(cG_{0}/2n)$, with $M $ a
nonnegative integer. In the case where the resonator gap is close to
a Bragg gap centered at the Brillouin zone edge, one can re-define
the reduced Brillouin zone so that the three mode approximation
remains a good approximation.
\end{references}
\end{document}